\documentclass[12pt,usenatbib]{article}
\usepackage{epsfig,amsmath,amssymb,natbib}
\usepackage{aas_macros}

\DeclareMathOperator{\sech}{sech}

\begin{document}

\title{A Gravitational Ising Model for the Statistical Bias of Galaxies}

\author{Andrew Repp\footnote{Corresponding author}\ \footnote{repp@ifa.hawaii.edu}\hspace{.1cm} \& Istv\'an Szapudi\footnote{szapudi@ifa.hawaii.edu}\\\small Institute for Astronomy, University of Hawaii, 2680 Woodlawn Drive, Honolulu, HI 96822, USA
\\
\\
\\
\small Essay written for the Gravity Research Foundation 2019 Awards for Essays on Gravitation
\\
\\
}

\date{\small\today}
\maketitle

\begin{abstract}
Evaluation of gravitational theories by means of cosmological data suffers from the fact that galaxies are biased tracers of dark matter. Current bias models focus primarily on high-density regions, whereas low-density regions carry significant amounts of information relevant to the constraint of dark energy and alternative gravity theories. Thus, proper treatment of both high and low densities is important for future surveys. Accordingly, we here present an interactionless Ising model for this bias, and we demonstrate that it exhibits a remarkably good fit to both Millennium Simulation and Sloan Digital Sky Survey data, at both density extremes. The quality of the fit indicates that galaxy formation is (to first order) an essentially local process determined by initial conditions.
\end{abstract}

\newpage
\section{Introduction}
\label{sec:intro}
Newton's deduction of the inverse square law from planetary orbits is perhaps the first use of astrophysics to constrain gravitational theories. Today we require a similar strategy: $\Lambda$CDM cosmology treats dark energy as a cosmological constant---one which our best theory overpredicts by $10^{120}$ \citep{Carroll2001}---whereas modified gravity mimics dark energy effects through other means. Hence, one goal of the \emph{Euclid} \citep{Euclid} and WFIRST \citep{WFIRST} surveys is to shed light on such questions using cosmological data.

These surveys, however, count galaxies rather than directly observing matter, and since galaxies are biased tracers of dark matter, the galaxy density is not proportional to the matter density. Thus, extracting cosmological information requires modeling of the relationship between the matter overdensity $\delta = \rho/\overline{\rho} - 1$  and the galaxy overdensity $\delta_g = N_\mathrm{gal}/\overline{N} - 1$ ($\rho$ being matter density and $N_\mathrm{gal}$ the galaxy counts in each three-dimensional survey pixel). The simplest bias models (e.g., \citealp{Hoffmann2017}) expand $\delta_g$ linearly ($\delta_g = b\delta$) or quadratically ($\delta_g = b_1 \delta + b_2\delta^2/2$), while a log bias (e.g., \citealp{delaTorre2013}) assumes $\ln(1+\delta_g) = b \ln(1+\delta)$. None of these phenomenological models perform particularly well compared to observations. Alternative approaches (e.g., halo models) are significantly more complex, and simulations require simplifications (of unknown impact) to render tractable the required range of scales.

Another drawback to these approaches is their focus on high-density regions, despite voids'  significant information content. In particular, low-density regions contain half of surveys' information on dark energy, since they constitute 50 percent of the initial volume of the universe (see also \citealp{WCS2015Forecast});  likewise, detection of modified gravity screening requires analysis of both ends of the density spectrum. Hence, we desire a bias model that is accurate at both extremes.

This essay presents such a model (Section~\ref{sec:model}). We demonstrate its accuracy compared to the Millennium Simulation and the Sloan Digital Sky Survey in Section~\ref{sec:acc}; we conclude in Section~\ref{sec:concl}.

\section{An Interactionless Ising Model}
\label{sec:model}
We begin with some physical assumptions. First, we assume the sites of galaxy formation are (to first order) determined by initial densities and local physics. Since $\ln(1 + \delta)$ is a unitless quantity roughly characterizing the initial conditions \citep{NSS09,CarronSzapudi2013}, we formulate our model in terms of $A\equiv\ln(1+\delta)$.
Second, we assume we can treat the subhalos in which galaxies form as roughly identical. We also note that each subhalo is in one of two states---either hosting a galaxy or not---and that the release of gravitational energy during galaxy formation renders hosting energetically favorable. Third, we assume that clustered galaxies occupy deeper potential wells than isolated galaxies, rendering galaxy formation more favorable in survey cells of higher overall density.

\begin{figure}
\leavevmode\epsfxsize=18cm\epsfbox{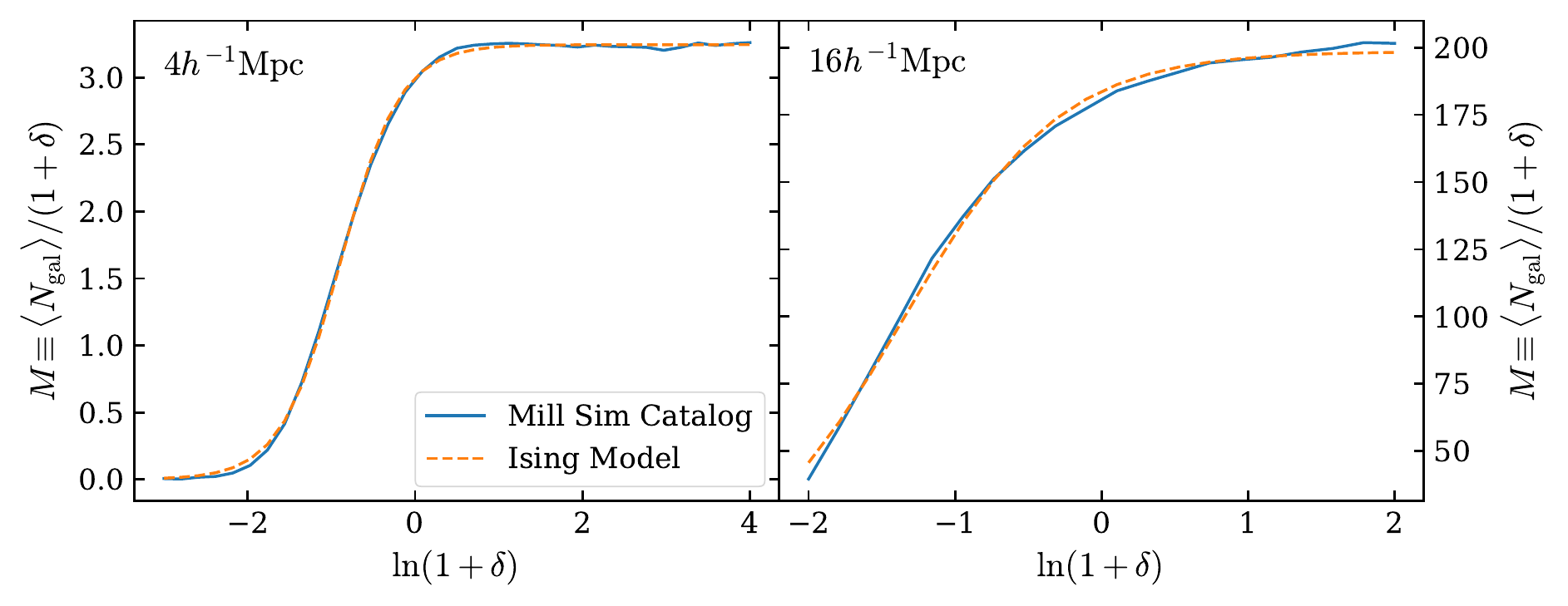}
\caption{Average number of galaxies per unit mass ($M$) as a function of $\ln(1+\delta)$, from Millennium Simulation galaxy catalogs (solid) and from fitting Equation~\ref{eq:final_FD} (dashed). The panels show two representative smoothing scales, namely, 4$h^{-1}$Mpc (left-hand panel) and 16$h^{-1}$Mpc (right-hand panel).}
\label{fig:MillSim}
\end{figure}

The Ising model nicely incorporates these assumptions. In this model, an atom has two possible states, one of which an external field renders energetically favorable; the model also typically allows interaction between neighboring atoms. Replacing the atoms with dark matter subhalos, we  thus consider an interactionless Ising model, which gives (e.g., \citealp{Pathria}) the following fraction of subhalos in a favorable (galaxy-hosting) state:
\begin{equation}
f_\mathrm{gal}  =  \frac{1}{2} e^{\beta E} \sech \beta E  =  \frac{1}{1 + e^{-2\beta E}},
\label{eq:Ising_no_J}
\end{equation}
where $E$ is energy and $\beta$  inverse temperature. After substituting a linear function of $A$ for (unitless) $\beta E$, Equation~\ref{eq:Ising_no_J} reduces to a Fermi-Dirac distribution:
\begin{equation}
f_\mathrm{gal} = \frac{1}{1 + \exp\left(\frac{A-A_t}{-T}\right)}\hspace{1cm}(T>0).
\label{eq:FD}
\end{equation}
Here $A_t$ marks the transition value between empty and filled subhalos. Note that subhalos in high-density cells fill up first because ``temperature'' ($-T$ in Equation~\ref{eq:FD}) is negative for gravitational collapse.

The Fermi-Dirac formulation suggests additional analogies: for instance, galaxies arguably obey an exclusion principle, since at most one galaxy can occupy a given subhalo. $A_t$ also corresponds to chemical potential, marking the density at which the next galaxy would form. Nevertheless, since galaxies cannot switch subhalos, we prefer to consider this model interactionless Ising rather than fermionic.

We now make the following observations: first, we wish to describe the expected number of galaxies per survey cell as a function of the underlying dark matter density. We thus assume that the expected number of subhalos in any pixel is proportional to the cell's matter density ($\langle N_\mathrm{sh} \rangle_A \propto 1 + \delta = e^A$). Thus we write
\begin{equation}
f_\mathrm{gal} = \frac{\langle N_\mathrm{gal} \rangle_A}{\langle N_\mathrm{sh}\rangle_A} = \frac{\langle N_\mathrm{gal} \rangle_A}{b\overline{N}(1 + \delta)},
\end{equation}
where $\overline{N}$ is the (global) mean number of galaxies per pixel. We can now write Equation~\ref{eq:FD} in terms of $M$, the expected number of galaxies per mass:
\begin{equation}
\label{eq:final_FD}
 M \equiv \langle N_\mathrm{gal} \rangle_A \cdot (1 + \delta)^{-1} = \frac{b\overline{N}}{1 + \exp\left(\frac{A_t-A}{T}\right)}\hspace{1cm}(T>0),
 \end{equation}
where, again, $A \equiv \ln(1+\delta)$.
In high-density regions ($A \gg A_t$), $M$ approaches $b\overline{N}$, so that $\langle N_\mathrm{gal} \rangle_A$ approaches $b\overline{N}(1+\delta)$: i.e., the number of galaxies is directly proportional to density. For low-density regions ($A \ll A_t$), the number of galaxies drops exponentially to zero.

Second, the matter power spectrum predominantly samples high-density regions, in which $(1+\delta_g) = \langle N_\mathrm{gal} \rangle_A/ \overline{N} = b(1+\delta)$, so that $\delta_g \sim b \delta$, as in the linear bias model. Thus a linear bias represents high-density regions fairly well but (as noted before) discards essential information on dark energy and modified gravity.

\begin{figure}
\leavevmode\epsfxsize=18cm\epsfbox{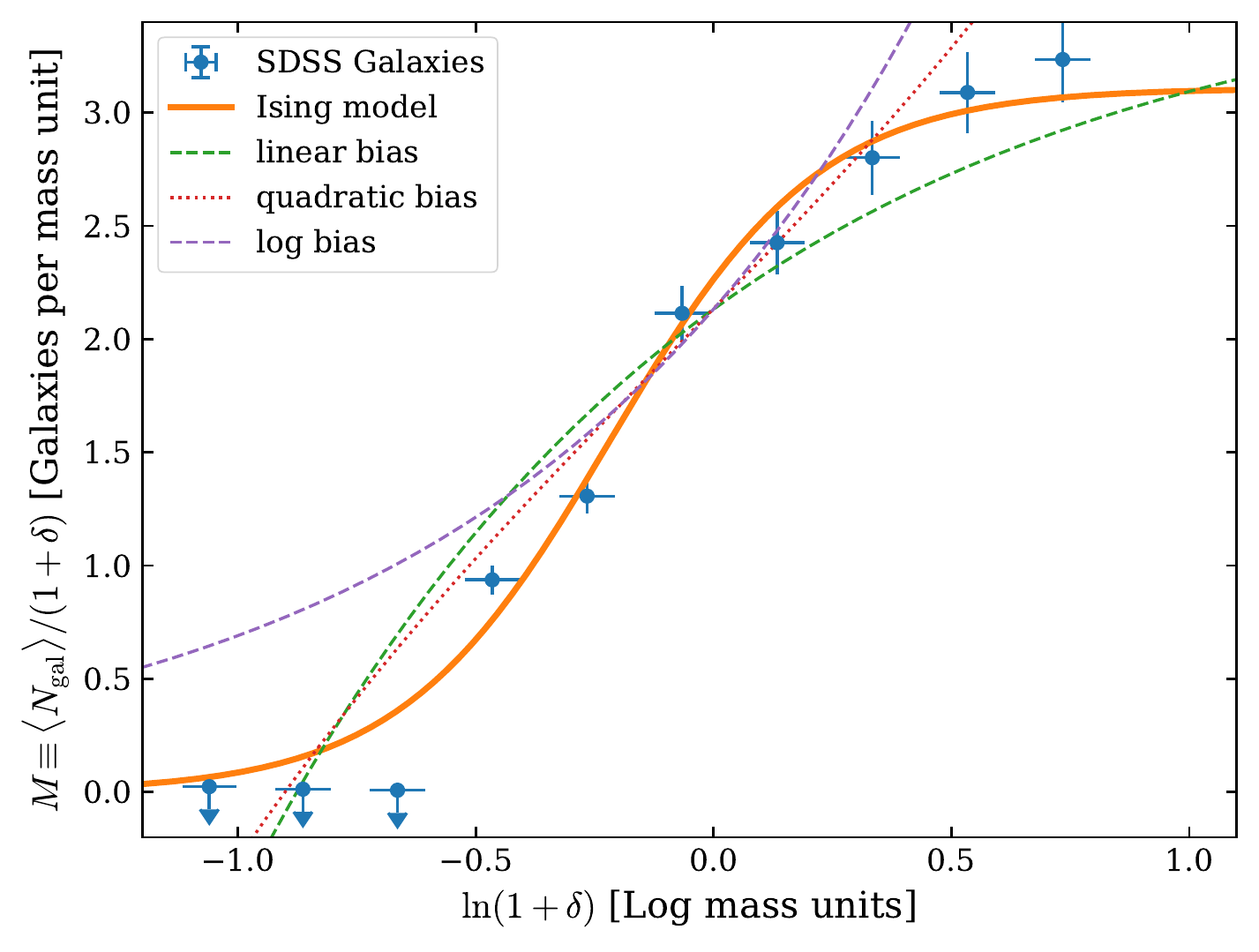}
\caption{Results from SDSS Main Galaxy Sample, compared to our model (Ising) and three other bias models. Horizontal error bars show the variation of $\ln(1+\delta)$ within each mass bin; vertical error bars reflect both the spread of measured $N_\mathrm{gal}$ values and the impact of the variation of $1+\delta$ within each bin. The three lowest bins contain no galaxies, and thus we show upper limits. Only the Ising model accurately reflects the galaxy distribution behavior for both high and low densities.
}
\label{fig:SDSS}
\end{figure}

Third, the various possible distributions of matter within a survey pixel produce a range of energetic favorabilities for cells of given (smoothed) density. The quantity $T$ parametrizes this effect of coarse-graining. On theoretical grounds we expect $T^2$ to be a decreasing function of log variance with a simple linear fit. Demonstrating this result is beyond the scope of this essay, but we here observe that a survey with larger pixels permits a greater variety of intrapixel matter distributions, and therefore $T$ increases with the survey pixel scale. (Just as $-T$ is analogous to temperature, the range of internal matter distributions is analogous to entropy.) For any given survey, however, $T$ depends on the survey parameters as well as the type of galaxy under consideration.

Fourth, the additional constraint provided by the total number of galaxies implies that this model has only two free parameters.

\section{Accuracy}
\label{sec:acc}

\subsection{Comparison to Simulation Data}
We first compare the model to galaxy catalogs obtained by applying semianalyical algorithms \citep{Croton2006a,DeLucia2006} to the Millennium Simulation \citep{Springel2005}. To test the impact of smoothing on various scales, we bin galaxies and dark matter into successively larger cubical pixels with side lengths 2, 4, 8, 16, and 32 $h^{-1}$Mpc; for each scale we calculate $M = \langle N_\mathrm{gal} \rangle_A \cdot (1+\delta)^{-1}$ as a function of $A=\ln(1+\delta)$. When we determine the best-fit parameters from Equation~\ref{eq:final_FD}, we find an excellent fit at all five scales. Figure~\ref{fig:MillSim} shows two examples.

\subsection{Comparison to Observational Data}
For comparison to observed galaxy distributions, we use a roughly volume-limited catalog from the Sloan Digital Sky Survey (SDSS) Main Galaxy Sample \citep{Blanton2005}. To achieve reasonable number densities, we bin galaxies into cubical cells with 16$h^{-1}$-Mpc sides, making cuts to avoid partially-filled pixels on sample borders.

We also need the underlying dark matter distribution; for this purpose we use the prescription of \citet{ReppApdf} at the mean redshift ($z = 0.15$), matching the quantiles of (observed) galaxy and (prescribed) matter distributions. Figure~\ref{fig:SDSS} shows the resulting values of $M$ alongside the best-fit Ising model, plus three other bias models.

The figure shows that the Ising model best captures the relationship between matter and galaxy densities. Of the others, the linear bias roughly reproduces the high-end plateau but quickly becomes unphysical ($N_\mathrm{gal} < 0$) at low densities. Indeed, at low densities the counts drop even faster than the Ising model predicts; nevertheless, our model is remarkably accurate given its lack of any halo modeling.
 
\section{Conclusions}
\label{sec:concl}
Constraining theories of gravity with cosmological data requires modeling the relationship between dark matter and galaxies, at both high and low densities. An interactionless Ising model (alternatively, a Fermi-Dirac model) yields a relationship that is both simple and surprisingly accurate.  Indeed, the success of this model provides fundamental insight into galaxy formation by validating our assumption that galaxy formation is (to first approximation) a local process, with tidal fields exerting only higher-order influence. Additionally, the fact that our independent variable $\ln(1+\delta)$ is a proxy for initial density strengthens our confidence that the initial conditions largely determine the galaxy sites.

Future work can extend this model with more accurate fits, perhaps by including interaction terms---just as \citet{ShethTormen} expanded on the earlier Press-Schechter (\citeyear{PressSchechter}) halo prescriptions. However, the model as it stands provides precisely what is necessary to place better constraints on dark energy and modified gravity.

\bibliographystyle{astron}
\bibliography{Thesis_Proposal}

\end{document}